\documentclass[prd,nofootinbib,superscriptaddress,twocolumn]{revtex4}
\usepackage{amsthm}

\usepackage{lmodern}
\usepackage[english]{babel}
\usepackage{amsmath}
\usepackage{amssymb}
\usepackage{mathptmx} 
\usepackage{color}
\usepackage{graphicx}
\usepackage{caption}
\usepackage{subcaption}
\usepackage{float}
\usepackage[all,cmtip]{xy}
\usepackage{bm}
\usepackage{enumerate}
\usepackage{pinlabel}

\usepackage{hyperref}

\usepackage{color,psfrag}
\usepackage{graphicx}
\usepackage{tikz}
\usetikzlibrary{calc}
\usetikzlibrary{decorations.pathmorphing}
\usetikzlibrary{shapes.geometric}
\usetikzlibrary{arrows,decorations.markings}
\newtheorem{theorem}{\bf Theorem}[section]

\begin{document}
\title{Totally Anti-symmetric Spinor Tensors in Minkowski Space}

\author{
	Peng Liu}

\address{College of Applied Science, Beijing Information Science $\&$ Technology University, 102206 Beijing, China}
\email{pliu@mail.ustc.edu.cn}

\author{Tanweer Sohail}

\address{Department of Mathematics, University of Jhang, Jhang, Pakistan} 

\email{tanveersms@gmail.com}

\author{Xiaoyu Jia}

\address{Beijing Institute of Education,  100032 Beijing, China}

\email{sherlyjxy2016@163.com}

\date{\today}

\begin{abstract}
	\baselineskip=16pt
	The spinor tensor $\epsilon_{AB}$ has a special property that its elements can be formulated into an algebraic expression of the indices. All the totally anti-symmetric tensors in Minkowski space are expressed by $\epsilon_{AB}$. By using the property, we give a simple proof of the total anti-symmetry for the volume spinor tensor.
\end{abstract}

\maketitle

\section{Introduction}
The application of spinors as a replacement for tensors to characterize physical quantities in four-dimensional Minkowski space is a profound achievement in theoretical physics, especially in the areas of electromagnetism, general relativity, and quantum field theory. The formalism exploits the internal attributes of spinors to provide conceptual simplicity as well as computational convenience. One of the most remarkable uses of spinor formalism is in the beautification of Maxwell's equations. Elegantly summarized as a single equation in the spinor formalism \cite{penrose1984spinors1}\cite{penrose1968twistor}\cite{penrose1969solutions}, Maxwell's equations are traditionally stated as a system of four coupled partial differential equations. This phenomenon lies in the fact that spinors possess an inherent capacity to represent the degrees of freedom of the electromagnetic field tensor, thereby enabling a more concise representation thereof. 

In addition, spinor formalism has been very useful in investigating knotted and linked configurations in electromagnetic fields \cite{bode2021stable}\cite{bialynicki2021new}. These topologically stable patterns, called electromagnetic knots, are solutions to Maxwell's equations and are of considerable importance in classical and quantum regimes \cite{liu2021yang}\cite{hu2015homfly}\cite{hu2016kauffman}. The spinor formulation of such knotted solutions is being ever more widely accepted as more fundamental than tensor formulations, insofar as they are more in harmony with spacetime's underlying  structure and offer a natural language for the further investigation.

Another compelling feature of the formalism is that the self-dual and anti-self-dual solutions to Maxwell's equations take on extremely simple form in terms of spinors. The duality is achieved by the contraction of the electromagnetic field tensor with a volume spinor tensor, an object that collects the four-dimensional spacetime volume's anti-symmetric properties. The volume spinor tensor, as defined in \cite{wald2010general}, is introduced by a concise formula acting as a foundation. However, the author of reference  \cite{wald2010general} does not provide a formal proof of its total anti-symmetry. Proving this anti-symmetric character is necessary for the spinor formalism to be consistent, since it is the basis for duality transformations and knotted solution construction to make sense.

In order to fill this lacuna, we advance a straightforward but strict proof of the full anti-symmetry of the volume spinor tensor. The approach is to cast the spinor tensor as an algebraic object specified by its indices, taking advantage of the structure of the spinor algebra in four dimensions. By taking the anti-symmetry down to the explicit calculation of this expression, we can show that the tensor actually does follow the required behavior under index permutations.

\section{Totally anti-symmetric spinor tensors}
In preparation for the formulation of the main theorem, we initially introduce the concept and notations of spinor by following \cite{wald2010general}. 

\begin{subsection}{The concept of spinor}
	
	For a two-dimensional vector space $\mathrm{W}$ over $\mathbb{C}$,  the anti-symmetric $(0,2;0,0)$ tensors, i.e. anti-symmetric bi-linear functions from $\mathrm{W}\times\mathrm{W}$ to $\mathbb{C}$, is one-dimensional. If such a non-degenerated tensor $\epsilon_{AB}=-\epsilon_{BA}$ is chosen, the pair $(\mathrm{W},\epsilon_{AB})$ is called a spinor space. The elements of $\mathrm{W}$ and the tensors over $\mathrm{W}$ are called spinors and spinor tensors. Any spinor $\xi^{A}$ in $\mathrm{W}$ can be mapped to the dual vector space $\mathrm{W}^*$ by $\epsilon_{AB}$, $\xi_{A}=\epsilon_{AB}\xi^{B}$. Since $\epsilon_{AB}$ is anti-symmetric, thus $$\xi_{A}\xi^{A}=\epsilon_{AB}\xi^{B}\xi^{A}=-\epsilon_{BA}\xi^{B}\xi^{A}=-\xi_{B}\xi^{B}=0.$$  
	If we consider the linear transformation $L$ on $\mathrm{W}$ keeping $\epsilon_{AB}$ invariant, $L$ should satisfy  $$\epsilon_{AB}L_{C}^{A}L_{D}^{B}=\epsilon_{CD},$$ 
	that is determinant $det L=1$ .
	Therefore, the group of $\epsilon_{AB}$ invariant linear transformations on $\mathrm{W}$ is exactly $SL(2,\mathbb{C})$.
	It is natural to use spinors to describe Minkowski space. Any point $(t,x,y,z)$ in the light cone in Minkowski space can be represented by a spinor $\xi^{A}$
	\begin{equation}\label{txyz}
		\left( \begin{array}{cc}
			t+z	& x+i y  \\
			x-i y	& t-z
		\end{array}\right) =\sqrt{2}\left(\begin{array}{c}
			\xi^{0}\\
			\xi^{1}
		\end{array} \right) \left(\begin{array}{cc}
			\bar{\xi}^{0'}& \bar{\xi}^{1'}
		\end{array} \right)  ,
	\end{equation}      
	where $\bar{\xi}^{A'}=\overline{\xi^{A}}$ is the complex conjugation of $\xi^{A}$.
	From the equation (\ref{txyz}), we can see that the light cone, i.e. $t^{2}-x^{2}-y^{2}-z^{2}=0$, is $SL(2,\mathbb{C})$ invariant. In fact, Minkowski metric $g_{ab}$ has a spinor expression $g_{AA'BB'}=\epsilon_{AB}\overline{\epsilon}_{A'B'}$. Thus, the equation (\ref*{txyz}) actually gives a map from  $SL(2,\mathbb{C})$ to Lorentz group. However this map is not one to one, the two elements in $SL(2,\mathbb{C})$ differed by plus and minus are mapped to one in the Lorentz group.  
	
	Minkowski space possesses a real dimension of four. Consequently, the totally anti-symmetric tensor space is a finite-dimensional real vector space. Subsequently, spinor expressions for the basis of these tensors will be provided. In the calculation, the utilization of the wedge product proves to be a convenient approach.  For any two spinors $\xi$ and $\eta$,  $\xi\wedge\eta$ is defined to be the anti-symmetric tensor product of $\xi$ and $\eta$:$$\xi\wedge\eta=\xi\otimes\eta-\eta\otimes\xi.$$ 
	If we choose the standard basis $\omicron^{A},\,\iota^{A} \in\mathrm{W}$ satisfying $\omicron_{A}\iota^{A}=1$, then the four real basis for Minkowski space are $\omicron^{A}\overline{\omicron}^{A'},\omicron^{A}\overline{\iota}^{A'}+\iota^{A}\overline{\omicron}^{A'},i(\omicron^{A}\overline{\iota}^{A'}-\iota^{A}\overline{\omicron}^{A'}),\iota^{A}\overline{\iota}^{A'}.$
\end{subsection}

\begin{subsection}{Totally anti-symmetric (0,2) tensors}
	The vector space of totally anti-symmetric real (0,2) tensors is six-dimensional, generated by the wedge product of four dual basis above. We can simplify them by using $\epsilon_{AB}=\omicron_{A}\iota_{B}-\iota_{A}\omicron_{B}$. For example, $$\omicron_{A}\overline{\omicron}_{A'}\wedge \iota_{B}\overline{\iota}_{B'}=\omicron_{A}\overline{\omicron}_{A'} \iota_{B}\overline{\iota}_{B'}- \iota_{A}\overline{\iota}_{A'}\omicron_{B}\overline{\omicron}_{B'}=\omicron_{A}\iota_{B}\overline{\epsilon}_{A'B'}+\epsilon_{AB}\overline{\iota}_{A'}\overline{\omicron}_{B'}.$$ For convenience, we omit the lower indices of $\omicron_{A},\iota_{A}$ and use $\omicron,\iota$ instead in the following.
	\begin{align*}
		&\omicron\overline{\omicron}\wedge(\omicron\overline{\iota}+\iota\overline{\omicron})=	\omicron\overline{\omicron}\omicron\overline{\iota}-\omicron\overline{\iota}\omicron\overline{\omicron}+\omicron\overline{\omicron}\iota\overline{\omicron}-\iota\overline{\omicron}\omicron\overline{\omicron}=\omicron\omicron\overline{\epsilon}+\epsilon\overline{\omicron}\overline{\omicron}\\
		&\omicron\overline{\omicron}\wedge i(\omicron\overline{\iota}-\iota\overline{\omicron})=i(	\omicron\overline{\omicron}\omicron\overline{\iota}-\omicron\overline{\iota}\omicron\overline{\omicron}-\omicron\overline{\omicron}\iota\overline{\omicron}+\iota\overline{\omicron}\omicron\overline{\omicron})=i(\omicron\omicron\overline{\epsilon}-\epsilon\overline{\omicron}\overline{\omicron})\\
		&(\omicron\overline{\iota}+\iota\overline{\omicron}) \wedge i(\omicron\overline{\iota}-\iota\overline{\omicron})=2i(\iota\overline{\omicron}\omicron\overline{\iota}-\omicron\overline{\iota}\iota\overline{\omicron})=2i(\iota\omicron\overline{\epsilon}-\epsilon\overline{\iota}\overline{\omicron})\\
		&\omicron\overline{\omicron}\wedge \iota\overline{\iota}=\omicron\overline{\omicron} \iota\overline{\iota}- \iota\overline{\iota}\omicron\overline{\omicron}=\omicron\iota\overline{\epsilon}+\epsilon\overline{\iota}\overline{\omicron}=\iota\omicron\overline{\epsilon}+\epsilon\overline{\omicron}\overline{\iota}\\
		&i(\omicron\overline{\iota}-\iota\overline{\omicron})\wedge \iota\overline{\iota}=i(\omicron\overline{\iota}\iota\overline{\iota}-\iota\overline{\omicron}\iota\overline{\iota}-\iota\overline{\iota}\omicron\overline{\iota}+\iota\overline{\iota}\iota\overline{\omicron})=i(\epsilon\overline{\iota}\overline{\iota}-\iota\iota\overline{\epsilon})\\
		&(\omicron\overline{\iota}+\iota\overline{\omicron})\wedge \iota\overline{\iota}=\omicron\overline{\iota}\iota\overline{\iota}+\iota\overline{\omicron}\iota\overline{\iota}-\iota\overline{\iota}\omicron\overline{\iota}-\iota\overline{\iota}\iota\overline{\omicron}=\epsilon\overline{\iota}\overline{\iota}+\iota\iota\overline{\epsilon}
	\end{align*}
	Thus, we have
	\begin{theorem}
		
		$\{\omicron\omicron\overline{\epsilon}+\epsilon\overline{\omicron}\overline{\omicron},i(\omicron\omicron\overline{\epsilon}-\epsilon\overline{\omicron}\overline{\omicron}),2i(\iota\omicron\overline{\epsilon}-\epsilon\overline{\iota}\overline{\omicron}),\iota\omicron\overline{\epsilon}+\epsilon\overline{\omicron}\overline{\iota},i(\epsilon\overline{\iota}\overline{\iota}-\iota\iota\overline{\epsilon}),\epsilon\overline{\iota}\overline{\iota}+\iota\iota\overline{\epsilon}\}$ forms a set of basis for the totally anti-symmetric (0,2) tensors of Minkowski space.  
	\end{theorem}

	Remark. We can see that all the basis of totally anti-symmetric (0,2) tensors are involved with six (0,2;0,2) tensors $\omicron\omicron\overline{\epsilon}$,$\iota\iota\overline{\epsilon}$,$(\iota\omicron+\omicron\iota)\overline{\epsilon}$,$\epsilon\overline{\omicron}\overline{\omicron}$,$\epsilon(\overline{\iota}\overline{\omicron}+\overline{\omicron}\overline{\iota})$,$\epsilon\overline{\iota}\overline{\iota}$.
	
\end{subsection}

\begin{subsection}{Totally anti-symmetric (0,3) tensors}
	
	Similarly, by the direct computation, the vector space of totally anti-symmetric real (0,3) tensors is four-dimensional, generated by
	\begin{align*}
		&-3\omicron\overline{\omicron}\wedge(\omicron\overline{\iota}+\iota\overline{\omicron})\wedge	\iota\overline{\iota}\\
		=&3[\omicron\iota\omicron(\overline{\omicron}\overline{\iota}\overline{\iota}-\overline{\iota}\overline{\iota}\overline{\omicron})+\omicron\omicron\iota(\overline{\iota}\overline{\omicron}\overline{\iota}-\overline{\omicron}\overline{\iota}\overline{\iota})-\iota\omicron\omicron(\overline{\iota}\overline{\omicron}\overline{\iota}-\overline{\iota}\overline{\iota}\overline{\omicron})]+c.c.\\
		=&\epsilon_{AB}\omicron_{C}(\overline{\epsilon}_{A'C'}\overline{\iota}_{B'}+\overline{\epsilon}_{B'C'}\overline{\iota}_{A'})+\epsilon_{AC}\omicron_{B}(\overline{\epsilon}_{B'A'}\overline{\iota}_{C'}+\overline{\epsilon}_{B'C'}\overline{\iota}_{A'})\\&+\epsilon_{BC}\omicron_{A}(\overline{\epsilon}_{B'A'}\overline{\iota}_{C'}+\overline{\epsilon}_{C'A'}\overline{\iota}_{B'})
		+c.c.\\
		&-3\omicron\overline{\omicron}\wedge i(\omicron\overline{\iota}-\iota\overline{\omicron})\wedge	 \iota\overline{\iota}\\
		=&3i[\omicron\iota\omicron(\overline{\omicron}\overline{\iota}\overline{\iota}-\overline{\iota}\overline{\iota}\overline{\omicron})+\omicron\omicron\iota(\overline{\iota}\overline{\omicron}\overline{\iota}-\overline{\omicron}\overline{\iota}\overline{\iota})-\iota\omicron\omicron(\overline{\iota}\overline{\omicron}\overline{\iota}-\overline{\iota}\overline{\iota}\overline{\omicron})]+c.c.\\
		=&i[\epsilon_{AB}\omicron_{C}(\overline{\epsilon}_{A'C'}\overline{\iota}_{B'}+\overline{\epsilon}_{B'C'}\overline{\iota}_{A'})+\epsilon_{AC}\omicron_{B}(\overline{\epsilon}_{B'A'}\overline{\iota}_{C'}+\overline{\epsilon}_{B'C'}\overline{\iota}_{A'})\\&+\epsilon_{BC}\omicron_{A}(\overline{\epsilon}_{B'A'}\overline{\iota}_{C'}+\overline{\epsilon}_{C'A'}\overline{\iota}_{B'})]+c.c.\\
		&\omicron\overline{\omicron}\wedge(\omicron\overline{\iota}+\iota\overline{\omicron}) \wedge i(\omicron\overline{\iota}-\iota\overline{\omicron})\\
		=&i[(\omicron\omicron\iota-\omicron\iota\omicron)\overline{\iota}\overline{\omicron}\overline{\omicron}+(\omicron\iota\omicron-\iota\omicron\omicron)\overline{\omicron}\overline{\omicron}\overline{\iota}-(\omicron\omicron\iota-\iota\omicron\omicron)\overline{\omicron}\overline{\iota}\overline{\omicron}]+c.c.\\
		=&2 i(\epsilon_{AB}\omicron_{C}\overline{\omicron}_{A'}\overline{\epsilon}_{B'C'}-\omicron_{A} \epsilon_{BC}\overline{\epsilon}_{A'B'} \overline{\omicron}_{C'})\\
		&(\omicron\overline{\iota}+\iota\overline{\omicron}) \wedge i(\omicron\overline{\iota}-\iota\overline{\omicron})\wedge\iota\overline{\iota}\;\\
		=&i[-\omicron\iota\iota(\overline{\iota}\overline{\omicron}\overline{\iota}-\overline{\iota}\overline{\iota}\overline{\omicron})-\iota\iota\omicron(\overline{\omicron}\overline{\iota}\overline{\iota}-\overline{\iota}\overline{\omicron}\overline{\iota})+\iota\omicron\iota(\overline{\omicron}\overline{\iota}\overline{\iota}-\overline{\iota}\overline{\iota}\overline{\omicron})]+c.c.\\
		=&2 i(\iota_{A}\epsilon_{BC}\overline{\epsilon}_{A'B'}\overline{\iota}_{C'}\;-\epsilon_{AB}\iota_{C}\overline{\iota}_{A'}\overline{\epsilon}_{B'C'}\;)\\
	\end{align*}
	where $c.c.$ is complex conjugate of the previous expression.
	Thus, we have
	\begin{theorem}
		\begin{align*}
		&\{\epsilon_{AB}\omicron_{C}(\overline{\epsilon}_{A'C'}\overline{\iota}_{B'}+\overline{\epsilon}_{B'C'}\overline{\iota}_{A'})+\epsilon_{AC}\omicron_{B}(\overline{\epsilon}_{B'A'}\overline{\iota}_{C'}+\overline{\epsilon}_{B'C'}\overline{\iota}_{A'})\\
		&+\epsilon_{BC}\omicron_{A}(\overline{\epsilon}_{B'A'}\overline{\iota}_{C'}+\overline{\epsilon}_{C'A'}\overline{\iota}_{B'})+c.c.,\\
		&i[\epsilon_{AB}\omicron_{C}(\overline{\epsilon}_{A'C'}\overline{\iota}_{B'}+\overline{\epsilon}_{B'C'}\overline{\iota}_{A'})+\epsilon_{AC}\omicron_{B}(\overline{\epsilon}_{B'A'}\overline{\iota}_{C'}+\overline{\epsilon}_{B'C'}\overline{\iota}_{A'})+\\
		&\epsilon_{BC}\omicron_{A}(\overline{\epsilon}_{B'A'}\overline{\iota}_{C'}+\overline{\epsilon}_{C'A'}\overline{\iota}_{B'})]
		+c.c.,\\
		&i(\epsilon_{AB}\omicron_{C}\overline{\omicron}_{A'}\overline{\epsilon}_{B'C'}-\omicron_{A} \epsilon_{BC}\overline{\epsilon}_{A'B'} \overline{\omicron}_{C'}),\\
		&i(\iota_{A}\epsilon_{BC}\overline{\epsilon}_{A'B'}\overline{\iota}_{C'}\;-\epsilon_{AB}\iota_{C}\overline{\iota}_{A'}\overline{\epsilon}_{B'C'})\}\\
		\end{align*} 
		forms a set of basis for the totally anti-symmetric (0,3) tensors of Minkowski space.  
	\end{theorem}
	
\end{subsection}

\begin{subsection}{Totally anti-symmetric (0,4) tensors}
	
	The totally anti-symmetric (0,4) tensors generates a real vector space of dimension one. The standard real basis $\mathrm{e}_{abcd}$, known as the volume form of Minkowski space \cite{wald2010general}, also possesses a spinor representation $$i \left( \epsilon_{AB}\epsilon_{CD}\bar{\epsilon}_{A'C'}\bar{\epsilon}_{B'D'}-\epsilon_{AC}\epsilon_{BD}\bar{\epsilon}_{A'B'}\bar{\epsilon}_{C'D'}\right).$$ We shall now present a simple proof for its total anti-symmetry.

	\begin{theorem}
		$$\mathrm{e}_{AA'BB'CC'DD'}=i \left( \epsilon_{AB}\epsilon_{CD}\bar{\epsilon}_{A'C'}\bar{\epsilon}_{B'D'}-\epsilon_{AC}\epsilon_{BD}\bar{\epsilon}_{A'B'}\bar{\epsilon}_{C'D'}\right)$$ is a totally anti-symmetric spinor tensor.
		
	\end{theorem}

	Proof. From the expression of $\mathrm{e}_{AA'BB'CC'DD'}$, it is obvious $\mathrm{e}_{AA'BB'CC'DD'}=\mathrm{e}_{CC'DD'AA'BB'}.$ Thus,  six equations of total anti-symmetry can be reduced to three: $$\mathrm{e}_{AA'BB'CC'DD'}+\mathrm{e}_{BB'AA'CC'DD'}=0,$$
	$$\mathrm{e}_{AA'BB'CC'DD'}+\mathrm{e}_{CC'BB'AA'DD'}=0,$$
	$$\mathrm{e}_{AA'BB'CC'DD'}+\mathrm{e}_{DD'BB'CC'AA'}=0.$$
	$\epsilon_{A B}$ is an anti-symmetric spinor tensor, where the index $A$ and $B$ is $0$ or $1$. It is clear that the value of $\epsilon_{AB}$ can be written as algebraic expression of its indices $\epsilon_{AB}=B-A$. Similarly,  $\bar{\epsilon}_{A'B'}=B'-A'$. Thus,   
$$\mathrm{e}_{AA'BB'CC'DD'}=i\left[\scalebox{0.8}{ (B-A)(D-C)(C'-A')(D'-B')-(C-A)(D-B)(B'-A')(D'-C')}\right],$$  
	$$\mathrm{e}_{BB'AA'CC'DD'}=i\left[\scalebox{0.8}{ (A-B)(D-C)(C'-B')(D'-A')-(C-B)(D-A)(A'-B')(D'-C')}\right],$$ 
	$$\mathrm{e}_{CC'BB'AA'DD'}=i\left[\scalebox{0.8}{ (B-C)(D-A)(A'-C')(D'-B')-(A-C)(D-B)(B'-C')(D'-A')}\right],$$ 
	\begin{align*}\mathrm{e}_{DD'BB'CC'AA'}=&i\left[\scalebox{0.8}{ (B-D)(A-C)(C'-D')(A'-B')-(C-D)(A-B)(B'-D')(A'-C')}\right].\\
		\end{align*}  
	Now it is straightforward to demonstrate the total anti-symmetry:
			\begin{align*}
		&\mathrm{e}_{AA'BB'CC'DD'}+\mathrm{e}_{BB'AA'CC'DD'}\\
		=& i\;[\; \scalebox{0.95}{ ( B\ -A\ )( D\ -C\ )\;(\; ( C'-A')( D'-B')-( C'-B')( D'-A')\;)\; }\\
		& -\scalebox{0.95}{( B'-A')( D'-C')\;(\; ( C\ -A\ )( D\ -B\ )-( C\ -B\ )( D\ -A\ ) \;)\;} ]\\
		=&i\; [\; \scalebox{0.95}{( B -A ) ( D -C)( C'-D')( A'-B')}\\& -\scalebox{0.95}{( B'-A')( D'-C')( C -D  ) ( A -B )} ]  	\\
		=&\scalebox{0.95}{ 0}. \\
		&\mathrm{e}_{AA'BB'CC'DD'}+\mathrm{e}_{CC'BB'AA'DD'}\\
		=& i\;[\; \scalebox{0.95}{ ( ( B\ -A\ )( D\ -C\ )- ( B\ -C\ )( D\ -A\ ))( C' -A' )( D' -B' )} \\
		& -\scalebox{0.95}{(( B' -A' )( D' -C' )- ( B' -C' )( D' -A' ))( C\ -A\ )( D\ -B\ )}]\\
		=& i\;[\; \scalebox{0.95}{( B -D )( A -C )( C' -A' )( D' -B' )}\\
		&-\scalebox{0.95}{( B' -D' )( A' -C' )( C -A )( D -B )}]  	\\
		=&\scalebox{0.95}{ 0}. \\
		&\mathrm{e}_{AA'BB'CC'DD'}+\mathrm{e}_{DD'BB'CC'AA'}\\
		=&i\;[\;\scalebox{0.95}{ (B-A)(D-C)(C'-A')(D'-B')-(C-A)(D-B)(B'-A')(D'-C')}\\
		&+\scalebox{0.95}{ (B-D)(A-C)(C'-D')(A'-B')-(C-D)(A-B)(B'-D')(A'-C')}\;]\;\\
		=&\scalebox{0.95}{ 0}.\\
	\end{align*}

	This completes the proof.	
	
\end{subsection}


\begin{section}{Discussion}
	Spinor description of physical quantities in Minkowski space, as illustrated in its application to Maxwell's equations, presents a powerful alternative to conventional tensor techniques. When the Maxwell equations were solved via the spinor formalism, the knotted solutions emerged spontaneously. The computational advantages of spinors are most clearly seen in numerical computation and symbolic manipulation, where the lower number of independent components results in more efficient algorithms. This efficiency is vital for simulating intricate systems, like those with knotted field configurations or self-dual solutions, that are computationally costly in tensor-based schemes. 
	
	The method in this paper relies on the property of the spinor tensor $\epsilon_{AB}=B-A$, which was a byproduct of studying the spinor description of knotted solutions. It was so fundamental yet had not been previously noted, and it might be expanded further reaching beyond electromagnetism into quantum computing, where the logic gates could be expressed by unitary operators acting on quantum states described by spinors. In addition, the consistency of the spinor formalism with the postulates of special relativity and its compatibility with the Dirac equation in quantum mechanics indicate its potential as a basis for future theoretical advancements. It also leaves open the possibility of investigating topological invariants of knots and links, or even more generally, quantum entanglement and the vacuum structure of quantum field theory.

\end{section}

	
\begin{section}{ACKNOWLEDGE}
	This article is dedicated to Professor Ke Wu in Capital Normal University in celebration of his 80th birthday.	
\end{section}	

\bibliography{reference}

\end{document}